# Neural Pathways of Responsible Gambling: How Personalized and Normative Messages Engage Gambling Severity and Individual Rationality

Juan Sánchez-Fernández , Luis-Alberto Casado-Aranda, İsmet Özer and Nuria-María Hernández-Vergara

**Abstract**

The way health recommendations are presented can shape how people perceive the information and its persuasion. This paper seeks to clarify whether the processing of two types of messages that promote responsible gambling, namely personalized messages (tailored suggestions based on gambling behaviors) and normative messages (which encourage behavior that aligns with socially acceptable standards), varies across gamblers' severity habits and rational thinking. Particularly, using functional Magnetic Resonance Imaging (fMRI), we examined brain activation patterns in response to both message types among 42 individuals with varying levels of gambling severity (assessed by the Problem Gambling Severity Index, PGSI) and self-perceived rationality (assessed by the trait prudence scale). Our findings support the hypothesis that personalized messages resonate more with individuals with higher PGSI scores, activating self-referential brain regions such as the medial prefrontal cortex more strongly. In contrast, normative messages are more effective for those with elevated self-perceived prudence scores, and they strongly engage mentalizing regions like the left angular gyrus. These results emphasize the role of personal traits in message receptivity and suggest that responsible gambling interventions may be more effective when the message type is aligned with an individual's severity and rationality profile.

# 1. Introduction

In 2023, more than 25% of the global population reported participating in gambling activities, with higher rates in countries like the USA (60%), the UK (54%), and Spain (49.5%) (Gambling Industry News, 2024). The rise of new technologies, combined with social pressures and the impact of advertising messages from operators (Griffiths et al., 2010), has contributed to an increase in gambling participation, particularly among individuals aged 18 to 30 (Derevensky & Gupta, 2007). Online gambling participation alone saw a 16% increase in participants worldwide, fueled by ease of access and 24/7 availability (Statista, 2024). This widespread engagement has escalated the prevalence of at-risk (8%) and problem gamblers (2%), who often face significant personal, financial, and social challenges due to compulsive gambling behaviors (Yan et al., 2016). Problematic gambling not only impacts individual well-being—leading to declines in mental health, self-esteem, and work productivity—but also places a significant economic burden on society, with estimated costs reaching billions annually (UK Goverment, 2023).

This growing public health issue calls for urgent intervention and effective strategies to mitigate the harmful effects of gambling. Responsible gambling (RG hereafter) initiatives involve developing tools that encourage players to engage in gambling activities with self-control, awareness, and an understanding of their personal limits. The goal is to ensure that gambling remains a safe and enjoyable form of entertainment, without resulting in negative financial or personal consequences (Hing et al., 2017). In recent years, a wide range of harm-minimization strategies have been developed to promote RG, with numerous studies demonstrating their effectiveness (Hing et al., 2018b). For example, self-imposed limits on time and spending have been widely demonstrated to help players manage impulsive behavior in gambling setting (Auer & Griffiths, 2015). Voluntary self-exclusion programs, which allow users to temporarily restrict access to gambling platforms, have proven effective in significantly reducing problem gambling behaviors (Hing et al., 2015; Gainsbury, 2014). Other measures, such as enforced breaks during play (Blaszczynski et al., 2016) and restrictions or modifications of note acceptors (Hansen & Rossow, 2010), have also shown positive outcomes. Additionally, online psychological support and brief cognitive-behavioral interventions have been found to meaningfully reduce gambling problems and enhance mental well-being (Dowling et al., 2017).

One harm-minimization initiative that has recently gained significant relevance to foster RG is health communication, that is, the use of communication evidence, strategy, theory, and creativity to advance the health and well-being of people and populations (Kim et al., 2013; Ray et al., 2024). Gambling studies have largely shown that health recommendations can play a significant role in promoting RG (Binde, 2014). For example, Hing et al. (2018) concluded that the use of emotive images, such as images of sadness or concern, is particularly effective in capturing the attention of young audiences and encouraging them to reflect on their gambling behavior. King et al. (2014) further highlighted that visual messaging can effectively capture adolescents' attention and encourage responsible behavior. Text-only messages, in turn, may be less visually engaging than graphic ads, but they can be effective in environments where gambling is perceived as a common activity, especially when direct and clear language is used to disrupt this perception. Color advertisements, especially those using alert colors such as red or yellow, are also effective in pop-ups or banners because these colors attract attention and convey urgency (Gainsbury & Blaszczynski, 2020). Health communication studies have also demonstrated that pop-up messages during gameplay are effective because they interrupt the activity, compelling players to pause and reflect (Hing et al., 2014).

A substantial body of research in health communication has focused on examining how different message frames (i.e., modes to present information) influence persuasion and gambling behaviors. For example, messages emphasizing potential gambling losses such as "How much are you willing to lose?" tend to evoke stronger emotional responses (vs. potential gains), potentially reducing impulsive gambling by promoting greater reflection (Parke & Parke, 2019). Kim et al. (2014) further suggested that time-focused messages, rather than financial ones, can encourage more reflective decision-making among players. Strong empirical evidence indicates that self-appraisal messages (i.e., formulating questions that prompt individuals to reflect on and evaluate their gambling behavior) are more effective than informational messages (i.e., direct instructions on how to engage in safer gambling) in reducing the amount of money wagered and gross gaming revenue (Gainsbury et al. 2015; Strecher 2007).

Two of the most widely investigated message frames in gambling studies are normative and personalized messages. Normative messages seek to shape an individual's gambling behavior by providing information about the behaviors, attitudes, or beliefs of a specific reference group, allowing gamblers to reflect on their actions in relation to those of their peers with similar characteristics (Auer & Griffiths, 2016). Personalized messages, in turn, provide personalized recommendations to promote RG based on an individual's existing beliefs, knowledge base, and gambling behaviors. Although both types of messages have been shown to be more persuasive than informative content (Auer & Griffiths, 2015), no consensus has been reached regarding which of these two message types is more persuasive. Lining with the rationale of Takahashi et al. (2023), we suggest that such lack of consensus may stem from four reasons. **First,** prior research often combines different message frames, such as personalized and normative ones, within a single message, which blurs their distinct effects. For instance, Auer and Griffiths (2015) examined how tailored information influenced gambling behavior. However, when designing the message, they also included data about comparable peers, which is concerning as it combined both personalized and normative information. **Second**, many studies prioritize real-world applicability, overlooking controlled factors such as word count and framing, which are essential for precise assessment. **Third**, while research on personality traits has examined how different characteristics of gamblers impact gambling behavior (Cunningham et al., 2012; Hing et al., 2016), no studies have yet explored how individual factors, such as gambling severity and rational thinking, interact with message characteristics to shape persuasion in promoting RG. **Fourth**, most health communication research relies on self-report measures, which are limited in capturing real-time cognitive and emotional responses. Those tools may need to be complemented by additional measurement methods, such as neuroimaging, that move beyond self-report limitations and help identify distinct neural correlates that unveil how personalized and normative messages are implicitly encoded and evaluated without requiring conscious participation of gamblers (Takahashi et al., 2023).

Aiming to overcome those limitations, this study explores the neural mechanisms underlying the processing of personalized and normative messages aimed at promoting RG, considering individual differences in personality and problematic gambling severity. Unlike traditional methods, neuroimaging has recently emerged as a highly promising approach in the field of communication (referred to as persuasion neuroscience) due to its ability to detect activity in deep brain structures that are inaccessible to other physiological techniques but vital to many emotional and cognitive processes influencing health-related decisions and behaviors (Donohew et al., 2018). By integrating findings from health communication research, persuasion neuroscience and gambling studies, we aim to provide a comprehensive framework that deepens our understanding of the persuasive power of personalized and normative messages based on their neural correlates. The following sections will explore the definitions and theoretical foundations of personalized and normative messages, how this persuasive power can be influenced by individuals' self-reported personal characteristics (gambling severity and individual capacity for making rational decisions), and the potential key role of neuroscience in assessing that persuasive effect.

## 2. Theoretical Background

### 2.1. Health communication promoting responsible gambling

#### 2.1.1. Personalized messaging

To combat noncommunicable diseases and promote healthier lifestyles, governments, healthcare institutions, and policymakers continue to develop strategies that encourage individuals to adopt healthier habits. Personalized messaging in health communication has emerged as a particularly effective intervention that provides personalized information aligned with a person's beliefs, attitudes, and behaviors (Kreuter et al., 2019). Tailored messages take into account a range of personal variables, such as baseline health behaviors, values, and motivations (Woolford et al., 2022), and address three key aspects: expected outcomes (e.g., health history and goals), efficacy expectations (benefits and barriers to healthier behaviors), and personalized elements, such as addressing the individual by name (Strecher et al., 2011). Research consistently shows that personalized messages are more persuasive than general informational approaches, especially in areas like nutrition, smoking cessation, and cardiovascular health. For instance, Casado-Aranda et al. (2021) reported that a tailored nutrition intervention reduced unhealthy eating, and Webb et al. (2020) found similar benefits for smoking cessation, with personalized messages perceived as more relevant and impactful.

In gambling research, technological advancements have enabled personalized messaging based on tracked gambling behaviors, which makes it possible to deliver personalized feedback via SMS or email. Personalized approaches in RG have shown promise by engaging the cognitive, emotional, and motivational aspects of behavioral change. For example, Monaghan & Blaszczynski (2010) demonstrated that personalized, reflective messages helped increase awareness about session lengths, resulting in higher likelihoods of breaks compared to generic warnings. Auer & Griffiths (2018) found that behavioral feedback tools utilizing individual real-time gambling data reduced gambling intensity among users. Along the same line, Auer and Griffiths (2020) found that online gamblers receiving personalized feedback (i.e., feedback concerning their own actual gambling behavior in the form of text messages) wagered significantly less money on both the day they read a personalized message and seven days after they read a personalized message. More recently, insights from Auer et al. (2023) on the Dutch Lottery suggested that players receiving personalized feedback made more withdrawals, supporting the idea that self-awareness through personalized messaging could help mitigate gambling risks and promote RG behaviors. However, the study by Hollingshead et al. (2020) concluded that personalized behavioral feedback did not enhance gamblers' adherence to limits.

The Elaboration Likelihood Model (ELM) by Petty & Cacioppo (1986) provides a strong psychological framework for understanding why personalized messaging is particularly persuasive in health domain. According to the ELM, individuals process information along two routes: the central and peripheral ones. When a message is personally relevant or aligned with an individual's beliefs and values, as in personalized messages, it is more likely to be processed via the central route, which involves careful, thoughtful evaluation of the information. This route enhances the likelihood of lasting attitude and behavioral change, because individuals engage deeply with the content, and compare it with their own experiences and values. In the context of RG, personalized messages that reflect individual gambling behaviors or risks may align stronger with the central route because they resonate with the user's personal experiences, making the message feel more relevant and urgent.

**2.1.2 Normative messaging**

Normative messaging has become an effective strategy for promoting healthy behaviors by encouraging individuals to align with socially acceptable or desirable norms. This approach leverages social norms—beliefs about what is commonly done or approved within a group—to influence health-related actions (Mollen et al., 2021). Studies on health communication have demonstrated that normative messaging can be more effective than simple informational approaches in raising intentions toward HPV vaccinations (Ratanasiripong, 2022), increasing mask-wearing during public health crises (Brooks et al., 2023), and promoting physical activity (Turner et al., 2021).

In the gambling behavior field, normative messaging could promote RG by aligning gambling habits with perceived social norms. For instance, Neighbors et al. (2015) conducted a computer-based intervention among college students with gambling issues, and found that normative feedback reduced perceived gambling norms, which subsequently led to decreases in gambling frequency and amount. Similarly, Berge et al. (2022) implemented a one-month normative intervention with young adult online casino users, demonstrating a notable reduction in gambling activity when participants were shown comparisons with peer behaviors. Additionally, Miller & Thomas (2018) evaluated pop-up messages during online gambling sessions, reminding users of the frequency with which others took breaks or set spending limits. These normative messages effectively increased the likelihood of players taking breaks and setting limits, especially among those susceptible to social influence. Eastwood et al. (2023) indeed found that normative feedback reduced young adults' gambling by reshaping their perceptions of normal gambling behaviors. The study by Browne & Rockloff (2021) showed that normative feedback about average gambling expenditures led to significant reductions in spending. However, some studies, such as Gainsbury & Blaszczynski (2020), have suggested that normative messages are only moderately effective unless paired with practical strategies, like budget tracking.

When comparing normative personalized message strategies, in a controlled experimental study, Auer and Griffiths (2016) tested three messaging approaches: personalized feedback, normative feedback, and simple recommendations. They observed that personalized messages significantly influenced gambling behavior, helping some players reduce their gambling time and spending. In contrast, normative feedback showed limited success in reducing gambling behavior. Supporting this, Auer and Griffiths (2015a), for example, examined the impact of normative and self-reflecting information in a real-world slot machine gambling environment by analyzing behavioral data from two representative random samples, each consisting of 800,000 gambling sessions (1.6 million sessions in total). Both types of messages—normative and self-related—were found to double the number of gamblers who stopped playing compared to simple recommendations. However, the authors were unable to determine which type of message was more effective in driving this behavioral change. Lining up with this reasoning, Harris et al., (2018) verified that the impact and the number of gamblers positively affected by both personalized and normative messages are limited.

The effectiveness of normative messaging can be analyzed through the lens of social psychology theories such as the Theory of Mind and Moral Cognition (Ho et al., 2022). Theory of Mind involves recognizing others' beliefs and emotions, which facilitates empathy and understanding of social expectations (Adams & Hansen, 2022). When individuals interpret normative messages, they engage in mentalizing—understanding others' perspectives and behavioral intentions—which fosters social alignment. Meanwhile, Moral Cognition involves evaluating actions within ethical and societal standards, prompting individuals to rationally judge their behaviors against social norms (Van Bavel et al., 2022).

## 2.2. Neuropsychological Mechanisms of personalized vs. normative messaging

Some of the aforementioned incongruities in the persuasion of personalized and normative messages may be attributed to the fact that most studies have relied on self-report methods, such as questionnaires, which do not adequately capture how gamblers process information cognitively and emotionally or track their reactions in real time. Indeed, because implicit, emotion-driven responses are central to the persuasion of gamble-related messages (Takahashi et al., 2023), additional tools are necessary to objectively measure these innermost underlying mechanisms beyond self-reporting. To meet these challenges, health communication scholars are lately borrowing theories and tools from biometrics and psychophysiology to evaluate the psychophysiological mechanisms underlying the nature and persuasion of health-related messages (so-called persuasion neuroscience) (Casado-Aranda et al., 2023). Relevant to our study, the research by Mutti-Packer et al. (2022) investigated attentional responses to messages highlighting the financial or social consequences of gambling. However, the authors found no significant differences in attention between socially and financially framed messages.

Despite these advances, no study to date has clarified the neural mechanisms underlying the processing of personalized and normative messages designed to promote RG, nor has any research compared their persuasive effects using neuroimaging data. This could constitute a step forward, as neuroimaging could enable a moment-by-moment measurement of physiological processes, localize activation to deep brain structures that cannot be accessed with other physiological techniques, spatially distinguish the neural bases of constructs like self-reference or mentalizing, and tackle unanswered questions in persuasion research without requiring gamblers to engage in conscious introspection

### 2.2.1 Neural correlates of personalized messages

According to the Elaboration Likelihood Model (ELM) by Petty and Cacioppo (1986), information that is personally relevant and motivational engages a central processing route, prompting individuals to critically evaluate the message against their own experiences and increasing the likelihood of behavioral change. Personalized messages, that often integrate self-relevance and motivation, are thus more likely to be deeply processed. Neuroscience research has extensively mapped the neural mechanisms involved in processing self-relevant information, highlighting the role of areas such as the mPFC and precuneus in fostering engagement with personalized content.

Along these lines, Chua et al. (2011) found that increased activation in the mPFC and precuneus when processing tailored messages for smoking cessation was linked to successful behavior change. Similarly, Falk et al. (2012) showed that mPFC activation during self-relevant health messaging could predict behavioral outcomes at a population level, while Peters and Büchel (2010) identified the mPFC's role in self-referential thinking and reward processing in personalized messages. Casado-Aranda et al. (2022) demonstrated that tailored nutritional messages activated self-relevance networks, including the precuneus and orbitofrontal cortex, while Vezich et al. (2016) found that alignment with personal beliefs and self-concept engaged the mPFC, ultimately promoting behavior change through increased sunscreen use.

### 2.2.2 Neural correlates of normative messages

Normative messages require the ability to understand and empathize with others' perspectives, as well as consider moral implications, and engage psychological mechanisms related to mentalizing and social cognition. Neuroimaging studies reveal the neural underpinnings of these processes, highlighting shared activation in regions like the temporo-parietal junction (TPJ), temporal lobes, angular gyrus, and dorsomedial prefrontal cortex (Dufour et al., 2013). Specifically, research underscores the role of mentalizing in social interactions, showing that social relevance is key to conformity. Cascio et al. (2015) found that adolescents who received feedback counter to their own evaluations showed increased activation in mentalizing regions, leading to a higher likelihood of conforming to group opinions. Similarly, Welborn et al. (2016) confirmed that mentalizing is critical for social influence, finding that conflicting feedback from parents and peers activated regions such as the vmPFC, TPJ, and angular gyrus, which correlate with increased susceptibility to peer influence.

Additionally, studies have indicated that normative messaging activates regions linked to social norms and empathy. Falk et al. (2012) demonstrated that the angular gyrus and superior frontal gyrus are engaged when participants are exposed to social norms, linking these regions to normative messaging responses. Seghier (2013) provided insights into the angular gyrus's role in perspective-taking and morality, essential for processing normative messages. Similarly, Decety & Cowell (2014) found that moral processing, often triggered by normative messages, activates areas like the angular and superior frontal gyrus, fostering empathy and norm processing. Van Overwalle & Baetens (2009) discussed the roles of these regions in the mentalizing system, while Chiao & Immordino-Yang (2013) emphasized that cultural and social norms stimulate areas like the angular and superior frontal gyrus, which are key for internalizing social norms. Collectively, these findings suggest that neural networks tied to mentalizing facilitate understanding others' perspectives and aligning personal views with social expectations.

## 2.3. Protection Motivation Theory (PMT): How individual characteristics affect the processing of gambling messages

Along with the limited use of biological tools to assess message persuasion in gambling contexts, the inconclusive findings on persuasion of gambling messages may stem from the lack of studies examining how gambler-related factors interact with message characteristics to shape persuasion. Fear theories, such as PMT (Rogers, 1975, 1983), provide a valuable framework for understanding how personal traits influence responses to messages, especially those related to risk and self-protection. Particularly, PMT proposes that individuals assess risk-related stimuli through two processes: threat appraisal and coping appraisal. Threat appraisal involves evaluating the perceived severity (how serious a threat is) and susceptibility (how likely one is to experience the threat) to the risk, while coping appraisal addresses the perceived effectiveness of a recommended action and one's self-efficacy (the ability to control one's gambling and follow the recommended behaviors).

On the one hand, the Problem Gambling Severity Index (PGSI) is one of the most widely used tools to measure the severity of gambling-related problems, and thus the individual's awareness of gambling-related potential consequences, a core factor in threat appraisal (Ferris & Wynne, 2001; Munoz et al., 2010). Originating from the Canadian Problem Gambling Index (Ferris & Wynne, 2001), the PGSI provides a standardized assessment of gambling behaviors across various contexts. Through a series of self-reported questions, it evaluates the frequency of risky gambling behaviors, their negative consequences, and their impact on daily life. It would be logical to report that the PGSI enables to identify individuals more likely to engage in self-reflective processing when exposed to messages that speak directly to their experiences and perceived gambling consequences (Northoff et al., 2011). In other words, high-PGSI gamblers have likely experienced significant gambling-related harms (financial loss, emotional distress, strained relationships). When encountering personalized messages about gambling risks, they can recognize their personal experiences in the message, and increase their perceived vulnerability and severity (when warranted). This direct connection between personal struggles and the message content could trigger heightened self-relevance.

The trait prudence scale complements the PGSI by capturing an individual's belief in their capacity for rational and prudent decision-making (Aguirre-Rodríguez & Torres, 2023). The self-perceived prudence is therefore crucial for understanding self-perceptions of cognitive processes, particularly in risk-laden contexts like gambling. Self-perceived rationality gages the degree to which individuals believe they make logical, methodical, and reasoned choices, rather than being swayed by emotions or impulsivity. Rational thinkers tend to evaluate messages through a logical framework, making them more receptive to messages that align with ethical and social norms, as Decety & Cowell (2014) suggested. Therefore, a self-perceived prudence is a useful indicator for identifying individuals who may respond more favorably to normative messages that emphasize social responsibility and cohesion, which aligns with PMT's coping appraisal.

## 3. The current research

Consequently, by integrating PMT with the Elaboration Likelihood Model (ELM) and the Theory of Mind and Moral Cognition, we can better grasp why messages tailored to personal vulnerabilities (PGSI) or moral and normative cues (self-perceived prudence) may elicit differing cognitive and emotional processing routes.

Particularly, individuals with higher PGSI scores tend to be more responsive to personalized messages, which often appeal to their personal gambling experiences and may engage self-referential cognitive processes. Because these messages are tailored to reflect the specific circumstances of each individual, they could be effective for those who are severely impacted by problem gambling, encouraging them to reflect on the personal consequences of their behavior. It would be logical, therefore, to propose that:

**RQ1:** Personalized messages will elicit stronger neural responses related to self-referential processing (particularly the mPFC and precuneus) in individuals with higher PGSI scores.

In contrast, individuals who perceive themselves as highly rational (high self-perceived prudence) are likely to find normative messaging more persuasive. Recent studies have suggested that individuals who perceive themselves as rational decision-makers tend to respond more favorably to messages that emphasize social responsibility and ethical behavior (Decety & Cowell, 2014). This is because rational individuals tend to process information through a logical and evaluative framework, making them more receptive to messages that encourage reflection on the broader societal and ethical consequences of their actions (Seghier, 2013). Therefore:

**RQ2:** Normative messages will elicit stronger neural responses related to mentalizing (such as the left and right angular gyrus, and superior frontal gyri) in individuals with high self-perceived prudence.

## 4. Methods

### 4.1. Participants

An online survey was circulated at a major university, resulting in a sample of 44 adult, right-handed, Spanish-speaking participants for an fMRI study. Data from two individuals were omitted as they did not complete the experiment's final tasks. To focus on individuals with low-to-moderate gambling risk, recruitment was guided by the PGSI scale (Ferris et al., 2017). The PGSI includes nine questions answered on a 4-point scale (never, sometimes, most of the time, almost always), covering three main areas: problematic gambling behavior (frequency and severity of risky actions, such as betting more than one can afford), negative consequences (the personal impacts of gambling, including financial issues, strained relationships, or guilt), and loss of control (the inability to control, reduce, or stop gambling behaviors). Participants were selected based on a PGSI score between 3 and 15: scores under 3 indicated no gambling issues, while scores between 3 and 7 suggested a potential for problem gambling. Scores between 7 and 15 indicated mild gambling problems, and scores above 15 denoted severe gambling issues. Rationality was measured using a 7-point Likert self-report scale that assessed factors such as self-control, foresight, responsibility, restraint, rationality, methodicalness, and planning (self-perceived trait prudence, Aguirre-Rodriguez & Torres, 2023). The descriptive statistics of the final simple are summarized in Table 1.

In addition, participants were excluded if they met any of the following criteria: (1) had mental health disorders such as schizophrenia, psychotic episodes, manic disorder, obsessive-compulsive disorder, alcohol or substance dependence, or post-traumatic stress disorder; (2) were undergoing mental health treatment unrelated to pathological gambling; (3) were under 18 years of age; (4) had a current or past brain injury; (5) were pregnant; or (6) were claustrophobic.

**Table 1.** Descriptive statistics of the sample

| Characteristics | Sample |
|---|---|
| N | 42 |
| Gender (F/M) | 1/41 |
| Age (years) | 22,9 ±4.7 |
| PSGI | 8.55±3.04 |
| Self-perceived prudence | 5.18±1.35 |

**4.2. Experimental procedure**

The main goal of the experimental setup was to present participants with three different types of messages intended to encourage RG: (i) personalized messages (e.g., "Mike, you keep your gambling hidden from your family because you're aware they'd be disappointed"), (ii) normative messages (e.g., "Most responsible gamblers play once a week or less. You can be part of this group"), and (iii) generic informational messages, which acted as a control for comparison (e.g., "Adrenaline and fear can make you lose track of time while gambling, impacting social skills"). The study followed a structured sequence: (1) an initial baseline session assessing participants' current gambling behaviors along with psychosocial, health, and demographic factors related to gambling addiction and RG behavior, used to create personalized messages; and (2) an fMRI session where participants viewed the three types of RG messages.

### 4.2.1. Session 1: Baseline session

The first session evaluated participants' behaviors, goals, characteristics, and challenges related to reducing irresponsible gambling through a video call interview. The collected responses were used to create personalized messages. Following frameworks by Latimer et al. (2005, 2010), Walthouwer et al. (2013), and Woolford et al. (2010), the focus was on several key variables: (i) gambling habits, where baseline questions explored participants' gambling frequency, weekly and monthly expenses, and intentions to cut down on these behaviors; (ii) outcome expectations, using open-ended questions to uncover individual support systems (e.g., family, friends, social environment) and goals around gambling and saving money; and (iii) efficacy expectations, which included attitudes and self-efficacy beliefs related to pathological gambling, such as participants' confidence in their ability to stop gambling. Additionally, the session assessed perceived risks (e.g., beliefs about whether gambling could harm or had already harmed their health) and barriers (e.g., stress, workload, academic obligations) as well as perceived benefits of RG behaviors.

### 4.2.2. Session 2: fMRI experiment

**Messages**

We crafted statements for each category of interest—personalized, normative, and informational messages. All three message types were structured similarly, maintaining a consistent word count (12-16). Based on prior studies, we ensured that personalized, normative, and informational messages covered comparable themes and balanced the inclusion of both loss- and gain-framed content. Personalized messages featured content tailored to each individual, incorporating responses from the baseline video interview. These messages reflected the participant's specific experiences and included positive (or negative) aspects of improving (or maintaining) their responsible (or irresponsible) gambling behaviors. Normative messages, on the other hand, were informed by general gambling behaviors among Spaniards, referencing the 2020 Annual Report on the Spanish Gambling Market and the 2022-2023 Gambling Prevalence Study from the General Directorate for the Regulation of Gambling. Lastly, informational messages, which were used as a control, provided general recommendations for achieving RG habits, based on guidelines from the JugarBien platform of the Spanish Ministry of Consumer Affairs.

**fMRI task**

One week after Session 1, the participants attended an fMRI session. During the scan, the participant was shown 20 personalized messages, 20 normative messages, and 20 informational messages. While the normative and informational messages were standardized across all participants, the personalized messages were customized according to each participant's responses from Session 1. Each message was displayed for 8 seconds, with a fixation period of 1 to 4 seconds before each, and the messages were presented in random sequence. The complete scan, including a 5-minute anatomical imaging segment, lasted approximately 18 minutes. E-Prime Professional 2.0 software was used to deliver the fMRI task.

**fMRI Analyses**

To identify the significantly activated brain regions for personalized and normative messages, we created personalized vs. informative and normative vs. informative contrasts. Furthermore, we confined the regions of interest to the association maps gathered from Neurosynth and applied them as inclusive masks when viewing the whole-brain results. The association map for the term "self-referential" was selected for the personalized vs informative contrast, while the "moral" term map was selected for the normative vs. informative contrast. The dependent samples t-test was applied to these contrasts to determine group level activation patterns among all participants in related areas. Additionally, cluster-level family-wise error rates (FWEc) were considered at the $p < 0.05$ significance threshold for small volume correction when considering significantly activated clusters.

Furthermore, we took each significant cluster as a ROI with activation related ß values extracted using the MarsBar toolbox. These parameter estimates were then correlated as predictors in in-sample regression models with a measure of PGSI and self-perceived prudence scores. For regression models predicting these items from ROI ß values, we employed robust regression models using the *lmrob* function from the robustbase R package. This function uses a MM-type regression model in which the data points are assigned weights based on their residuals, as described in Yohai (1987) and Koller & Stahel (2011).

**5. Results**

While raw T1 and T2* weighted MRI scans can be found at https://openneuro.org/datasets/ds005357, non-thresholded statistical maps of the related contrasts are again openly available at https://neurovault.org/collections/17641/. The code can be consulted in https://osf.io/2x4pf/. The whole brain results of the personalized (vs informative) message contrast with the inclusive mask of "self-referential" systems map gave out 4 significantly activated clusters (Table 2).

**Table 2.** Whole brain results of the personalized > informative message contrast with the inclusive mask of "self-referential" Neurosynth association map (FWEc p < 0.05; coordinates are given in MNI space).

| Region | Hemi | n voxels | T | x | y | z |
|---|---|---|---|---|---|---|
| Medial Prefrontal Cortex (mPFC) | L | 152 | 15.72 | -2 | 52 | 10 |
| | R | | 14.18 | 1 | 60 | 14 |
| | R | | 12.16 | 8 | 56 | 21 |
| | R | | 9.95 | 8 | 49 | 32 |
| | L | | 7.92 | -13 | 60 | 21 |
| | R | | 7.03 | 8 | 56 | -4 |
| Precuneus | L | 78 | 14.54 | -2 | -56 | 32 |
| | R | | 12.82 | 4 | -52 | 28 |
| | R | | 12.38 | 1 | -63 | 32 |
| Superior Frontal Gyrus | L | 37 | 12.06 | -6 | 52 | 28 |
| | R | | 5.69 | 1 | 38 | 28 |
| Angular Gyrus | L | 32 | 11.97 | -52 | -60 | 24 |
| | L | | 9.31 | -44 | -66 | 38 |
| | L | | 6.52 | -38 | -74 | 38 |

Further robust regression analysis indicated a significant prediction of PGSI scores from the β values of the mPFC. The model demonstrated a coefficient estimate of 6.22 (SE = 2.43), with a 95% confidence interval ranging from 1.31 to 11.13. The predictor was statistically significant, $t(39) = 2.56$, $p = 0.014$. The model explained 13.4% of the variance in PGSI.

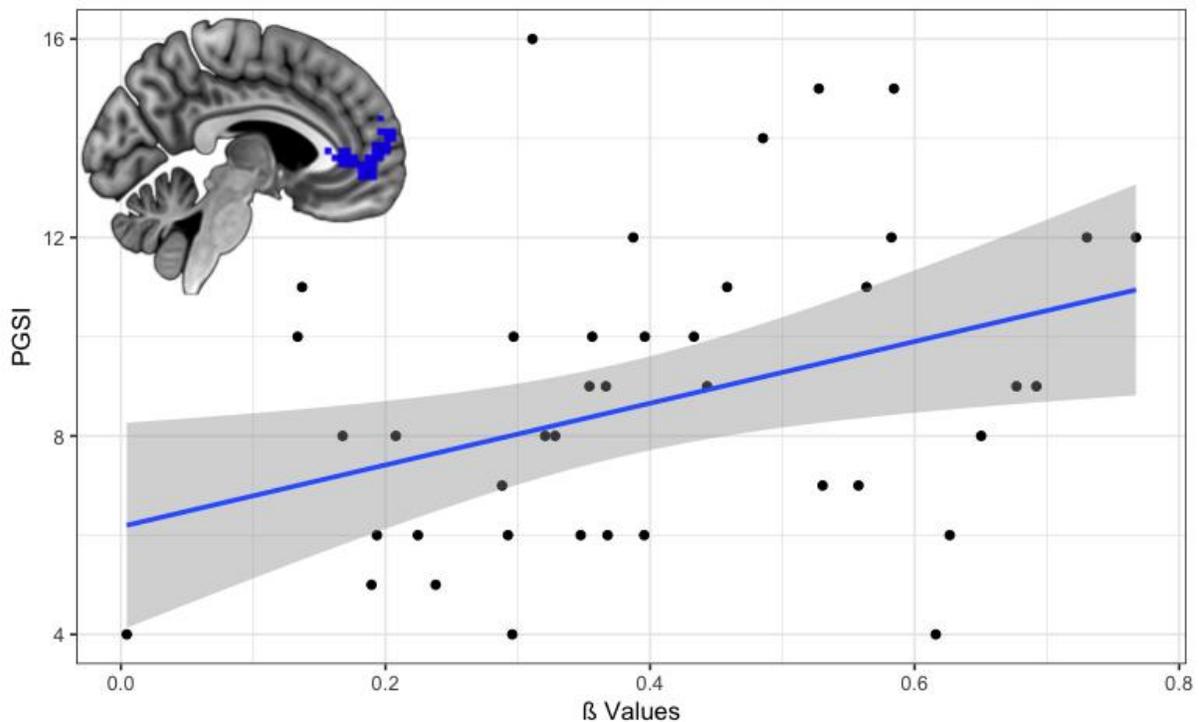

**Figure 1.** Robust regression line predicting PGSI from the ß values of the medial prefrontal cortex activation cluster ROI extracted from the personalized > informative contrast with the "self-referential" Neurosynth map (x = -4).

The whole brain results of the normative (vs informative) message contrast with the inclusive mask of "moral" systems map showed 3 significantly activated clusters (Table 3).

**Table 3**. Whole brain results of the normative > informative message contrast with the inclusive mask of "moral" Neurosynth association map (FWEc p < 0.05; coordinates are given in MNI space).

| Region | Hemi | n voxels | T | x | y | z |
| --- | --- | --- | --- | --- | --- | --- |
| Angular Gyrus | R | 33 | 9.96 | 50 | -56 | 28 |
| Angular Gyrus | L | 32 | 8.19 | -52 | -60 | 28 |
| Superior Frontal Gyrus | R | 38 | 6.42 | 4 | 49 | 28 |
|  | L |  | 5.58 | -6 | 46 | 35 |
|  | R |  | 5.06 | 8 | 56 | 32 |

Robust regression analysis indicated a significant prediction of subjective rationality (self-perceived prudence) scores from the β values of the left Angular Gyrus. The model demonstrated a coefficient estimate of 1.73 (SE = 0.70), with a 95% confidence interval ranging from 0.32 to 3.14. The predictor was statistically significant, $t(39) = 2.48$, $p = 0.018$. The model explained 10% of the variance in subjective rationality.

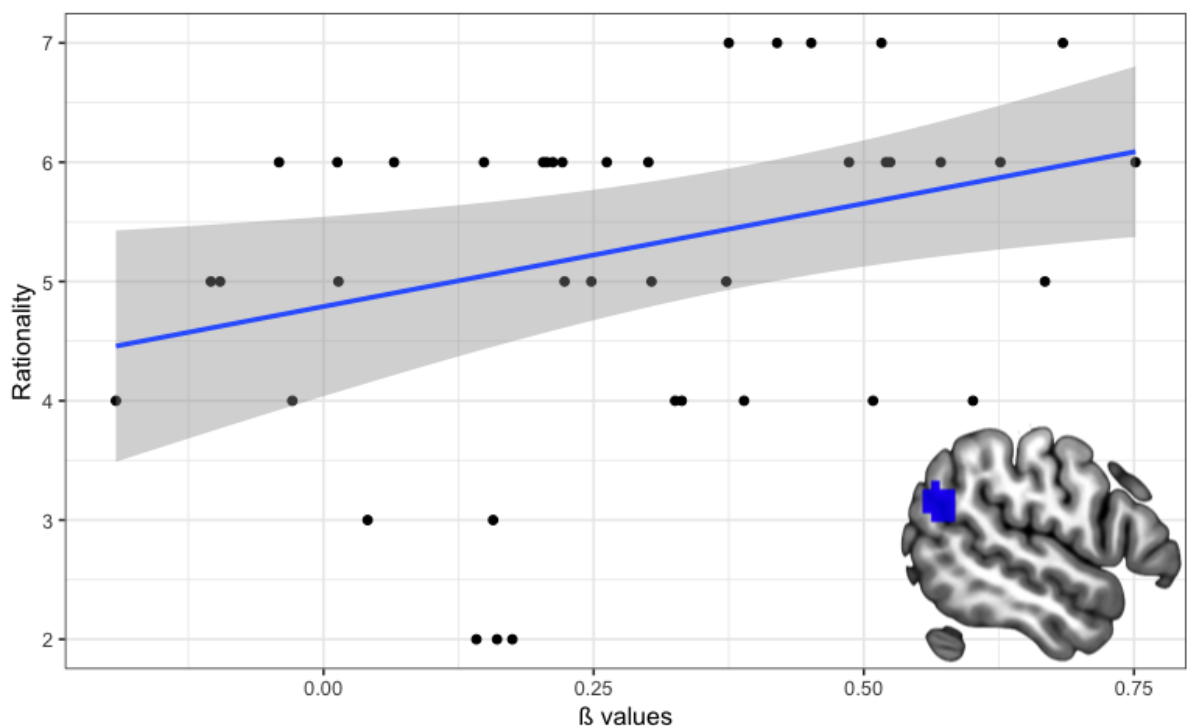

**Figure 2**. Robust regression line predicting subjective rationality from the ß values of left Angular Gyrus activation cluster ROI extracted from the normative > informative contrast with the "moral" Neurosynth map (x = -52).

To account for alternative predictions, we also fitted robust regression analyses trying to predict morality scores from ROI parameter estimates of the significantly activated clusters within the self-referential associations map in the Self vs informative contrast. Robust regression analysis indicated a non-significant prediction of rationality scores from the β values of the Left Medial Prefrontal Cortex. The model demonstrated a coefficient estimate of -0.28 (SE = 1.17), with a 95% confidence interval ranging from -2.64 to 2.08. The predictor was not statistically significant, $t(39) = -0.24$, $p = 0.812$. Similarly, robust regression analysis indicated a non-significant prediction of rationality scores from the β values of the Left Precuneus. The model demonstrated a coefficient estimate of 0.52 (SE = 0.75), with a 95% confidence interval ranging from -0.99 to 2.03. The predictor was not statistically significant, $t(39) = 0.69$, $p = 0.493$. Along the same line, robust regression analysis indicated a non-significant prediction of rationality scores from the β values of the Left Superior Frontal Gyrus. The model demonstrated a coefficient estimate of 0.23 (SE = 1.10), with a 95% confidence interval ranging from -2.00 to 2.45. The predictor was not statistically significant, $t(39) = 0.20$, $p = 0.839$. Lining up with those findings, robust regression analysis indicated a non-significant prediction of rationality l scores from the β values of the Left Angular Gyrus. The model demonstrated a coefficient estimate of 0.26 (SE = 1.13), with a 95% confidence interval ranging from -2.01 to 2.54. The predictor was not statistically significant, $t(39) = 0.23$, $p = 0.816$.

Furthermore, we fitted robust regression analyses trying to predict PGSI from ROI parameter estimates of the significantly activated clusters within the moral associations map in the Normative vs informative contrast. The robust regression analysis indicated a non-significant prediction of PGSI scores from the β values of the Right Angular Gyrus. The model demonstrated a coefficient estimate of 0.48 (SE = 1.68), with a 95% confidence interval ranging from -2.91 to 3.87. The predictor was not statistically significant, $t(39) = 0.29$, $p = 0.776$. Along the same line, robust regression analysis indicated a non-significant prediction of PGSI scores from the β values of the Left Angular Gyrus. The model demonstrated a coefficient estimate of 2.70 (SE = 2.31), with a 95% confidence interval ranging from -1.97 to 7.37. The predictor was not statistically significant, $t(39) = 1.17$, $p = 0.249$. Similarly, robust regression analysis indicated a non-significant prediction of PGSI scores from the β values of the Right Superior Frontal Gyrus. The model demonstrated a coefficient estimate of 3.32 (SE = 3.83), with a 95% confidence interval ranging from -4.44 to 11.07. The predictor was not statistically significant, $t(39) = 0.87$, $p = 0.392$.

The unprocessed T1 and T2* weighted scans can be found at [https://openneuro.org/datasets/ds005357](https://openneuro.org/datasets/ds005357), and the small volume corrected statistical maps of the personalized > informative (within the "self-referential" inclusive mask) and normative > informative (within the "moral" inclusive mask) contrast clusters can be found at [https://neurovault.org/collections/18157](https://neurovault.org/collections/18157).

## 6. Discussion

As with all human behaviors, gambling is complex and influenced by various facilitators, psychological factors, and external influences that shape individual motivations. In this study, we examined how two distinct communication styles—personalized and normative—resonate differently among gamblers with varying personality traits. We refer to personalized messages as those tailored to an individual's attitudes, preferences, and gambling habits, whereas normative messages promote behaviors that align with socially accepted gambling standards. In an exploratory approach, we employed functional Magnetic Resonance Imaging (fMRI) for the first time to investigate how these message types are processed in the brain, considering individual differences in rationality and problematic gambling severity. Our findings reveal that personalized messages resonate more with individuals exhibiting higher gambling severity, eliciting stronger activation in self-referential brain regions such as the medial prefrontal cortex. In contrast, normative messages are more effective for individuals with higher rationality scores, engaging brain areas associated with mentalizing, such as the left angular gyrus.

More specifically, the results confirm **RQ1**, which proposed that personalized messages more strongly engage brain areas associated with self-referential processing in gamblers reporting high severity scores. In particular, we demonstrate a significant relationship between mPFC activation and PGSI scores when participants read personalized messages. The mPFC is a key region involved in self-referential processing, and has been shown to engage when individuals consider personal and autobiographical information (Northoff et al., 2011; Yassin et al., 2022). This suggests that personalized messages, that emphasize self-relevant content, encourage individuals to reflect on their personal experiences with gambling as observed in brain areas associated with self-reference. For individuals with higher PGSI scores, who are more prone to gambling-related issues, this self-referential activation may prompt them to think more critically about how gambling affects their lives. Prior research has revealed that self-referential thinking plays a critical role in influencing behavior, particularly in individuals who are more susceptible to problematic health behaviors (Casado-Aranda et al., 2022; Chua et al., 2009; Jenkins & Mitchell, 2011). Thus, the findings support the idea that personalized messages resonate more with individuals at higher risk by making the consequences of gambling personally relevant, potentially motivating behavior change by increasing self-awareness of the risks involved (Mezulis et al., 2004).

**RQ2** anticipated that normative messages would elicit stronger activation in mentalizing-related brain areas, particularly in gamblers who perceive themselves as more rational, due to their greater ability to follow socially accepted recommendations. Our findings confirmed this, revealing a significant relationship between left angular gyrus activation while reading normative messages and rationality scores. Previous studies have identified the angular gyrus as central to moral reasoning, perspective-taking, and social cognition (Bzdok et al., 2012; Seghier, 2013). Its activation suggests that normative messages, which emphasize societal and moral implications, engage brain regions involved in higher-order reasoning and ethical judgment—particularly in self-perceived rational individuals, who may be more receptive to such messages as they encourage a logical evaluation of gambling behavior within a social framework. This supports the idea that normative messages appeal to mentalizing and rational decision-making, prompting gamblers to reflect on their behavior through a moral lens. Individuals with heightened rationality may respond more positively to these messages, seeing them as consistent with their values of self-regulation and social responsibility (Decety & Cowell, 2014; Saxe & Kanwisher, 2003).

Theoretically, our findings represent a fourfold contribution. **First**, the distinct brain activations observed during the processing of personalized and normative messages suggest that these messages influence gambling behavior through different neural pathways—engaging self-referential processing and rational moral reasoning, respectively. By targeting separate cognitive mechanisms, these messages can complement each other in shaping gambling-related decision-making. Beyond a rational assessment of gambling habits, mPFC activation in response to personalized messages may also reflect an emotional component, as this region integrates affective states with self-referential cognition (D'Argembeau et al., 2005; Northoff et al., 2011). Emotionally resonant content could enhance message salience and encourage deeper reflection, increasing the likelihood of behavior change through heightened personal relevance. In contrast, normative messages may engage social reward mechanisms by activating regions involved in understanding others' beliefs and moral reasoning, such as the angular gyrus and temporoparietal junction—areas linked to empathy, social values, and moral evaluation (Decety & Cowell, 2014; Moll, de Oliveira-Souza, & Zahn, 2008). These messages could shape behavior through anticipated social approval or disapproval, rather than internal reflection.

**Second**, this study makes a significant theoretical contribution by bridging neuroscience, behavior change, and health communication domains, demonstrating that message efficacy is not only a function of content but is also deeply influenced by individual characteristics of the message recipient. The integration of neuroimaging with health communication provides a nuanced understanding of how tailored messages interact with the brain systems responsible for self-referential processing and social cognition. Unlike previous research that broadly classified messages as personalized or normative, this study sheds light on the specific neural pathways activated by each message type based on recipients' personal characteristics. In this way, our study provides the first evidence of neural processing differences between two message types, whose persuasive impact has yielded inconclusive results in the gambling behavior literature (Auer and Griffiths (2015); Auer and Griffiths, 2016; Harris et al., 2018).

**Third,** by identifying distinct neural correlates associated with self-referential processing (mPFC) for personalized messages and mentalizing processes (left angular gyrus) for normative messages, this study advances theoretical models on persuasive communication, and shows that personalization engages cognitive and emotional dimensions differently than normative messaging. These findings offer insights into the Elaboration Likelihood Model (ELM), specifically suggesting that message relevance—whether derived from self-relevance or social norms—can activate central processing routes that enhance engagement and potentially lead to behavioral change (Petty & Cacioppo, 1986). This study also underscores the role of individual factors, such as gambling severity and perceived rationality, in moderating message persuasive power, thus providing empirical support for a more dynamic, recipient-oriented model of health persuasion. By further considerations of the findings in light of theoretical frameworks suggested in the introduction, a more comprehensive interpretation of the observed neural patterns can be seen. PMT, for instance, suggests that individuals' responses to health-related issues depend on their appraisal of severity and vulnerability, as well as their perceived effectiveness to cope (Milne, Sheeran, & Orbell, 2000; Rogers, 1983). In the context of this study, individuals with higher PGSI scores, who likely perceive greater severity, may find personalized messages more persuasive because these messages relate to personally relevant risks, thereby enhancing the threat perception and prompting more engaged processing. Similarly, from a moral cognition perspective, norms-based messages recruit brain regions linked to understanding others' mental states and values, aligning with frameworks that emphasize the role of empathy and moral judgment in guiding behavior (Decety & Cowell, 2014). By mapping these results onto PMT and moral cognition models, we gain insight into how harm severity and rationality may align with distinct motivational and coping strategies. While the ELM provides a valuable structure, integrating PMT's emphasis on threat and coping and moral cognition's focus on social and ethical standards can help explaining why certain subgroups, defined by their gambling severity or rational self-perception, may resonate more to differing message formats.

Lastly, our study makes a headway in the persuasion neuroscience literature, by revealing the neural bases of two message types scarcely explored so far. Prior studies identified the neural patterns of messages focused on how vs. why (Vezich et al., 2016), hedonic vs. utilitarian (Casado-Aranda et al., 2022) or future vs. past (Casado-Aranda et al., 2018). Our study not only extend the neural patterns observed in message tailoring within smoking (Chua et al., 2009) and nutritional contexts (Casado-Aranda et al., 2022) to the gambling domain; but sheds light by first time to the neural processing of normative messages.

The practical implications of this study are substantial for stakeholders involved in promoting RG. Regulatory bodies and policymakers can use these insights to design interventions that maximize impact by segmenting audiences based on individual risk profiles and cognitive characteristics. For instance, individuals with high PGSI scores may benefit from interventions that incorporate personalized feedback, directly addressing their specific habits and patterns, whereas those with a strong orientation toward rational decision-making may respond better to normative messaging that aligns with social and ethical standards. Mental health professionals and behavioral therapists can apply these insights to enhance therapeutic interventions, using tailored messaging to support clients in their journey toward healthier gambling behaviors. Personalized messages could be incorporated into cognitive-behavioral therapy (CBT) frameworks to facilitate self-reflection, while normative messages could be used to reinforce social accountability and community support. Finally, advocacy groups and social organizations dedicated to gambling prevention can use these insights to design targeted campaigns that resonate with diverse audiences. The development of materials that either focus on personal reflection or emphasize social norms, can foster greater public awareness and self-regulation among at-risk individuals.

## 7. Limitations and future research

This study offers valuable insights, but it has several limitations. Although the fMRI-based approach is effective in identifying brain activation patterns, is limited in capturing the long-term effects of these messages on behavior change. The use of self-reported scales for PGSI and self-perceived prudence introduces potential biases, because these measures rely on subjective perceptions that may not fully capture the underlying psychological factors. Moreover, the study did not account for the influence of cultural and environmental variables, which could shape individuals' responses to gambling messages. Different sociocultural contexts might alter the perception and processing of normative and personalized messages, indicating the need for cross-cultural research to understand these dynamics.

The related fields can benefit from future studies that take a more in depth look at how both personalized messages as well as normative messages interact with cognitive biases, metacognition, and heuristic reasoning to further affect the decisions of gamblers. People who suffer from the gambler's fallacy or confirmation bias, for example, may react differently when given personalized feedback about their specific betting results because it threatens their false sense of control or predictability of winning streaks (Toneatto, 1999; Goodie & Fortune, 2013). Normative messages may highlight what is normal or problematic to counter the effects of memorable but infrequent wins. Perhaps players can be steered toward more realistic expectations that are more in line with collective standards (Griffiths, 1994; Hodgins et al., 2011). Further, a gambler's ability to recognise and control their own understanding and thinking ability may play a role in how they internalize the gambler intervention. The personalized messages could get people to think critically about their judgement and reasoning patterns, enhancing more deliberative self-reflection of flawed strategies. On the other hand, normative messages could bolster socially supported norms which discourage impulsive heuristic-based decision-making. As strengthening metacognitive skills can assist individuals in maintaining healthier gambling behaviors that not only persist after the first exposure to tailored messages, but also enhance their own self-awareness and sustainable risk management (Zimmerman, 2000). Additionally, future research should consider a longitudinal approach to examine how repeated exposure to personalized and normative messages influences gambling behavior over time. Furthermore, integrating context-specific factors, such as cultural background and socioeconomic status, could provide a more comprehensive understanding of how diverse populations respond to RG messages. Exploring the combination of tailored messages with complementary behavioral interventions, such as digital tools for self-monitoring or goal-setting, could also yield insights into multifaceted approaches for promoting RG. Furthermore, advancements in neuroimaging techniques may allow for the exploration of other neural networks involved in emotional regulation and impulse control, offering a richer perspective on how tailored messages interact with various aspects of cognitive and emotional functioning. Finally, examining the role of emerging media formats, such as virtual reality and interactive applications, could open new avenues for delivering tailored interventions that are engaging, immersive, and impactful.

Auer, M., & Griffiths, M. D. (2015). Testing normative and self-appraisal feedback in an online slot-machine pop-up in a real-world setting. *Frontiers in Psychology*, *6*, 339.

Auer, M., & Griffiths, M. D. (2018). Cognitive behavioral therapy for gambling disorder: Examining efficacy using real gambling data and behavioral feedback. *Journal of Behavioral Addictions*, *7*(1), 28-36. https://doi.org/10.1556/2006.7.2018.01

Auer, M., & Griffiths, M. D. (2020). The use of personalized messages on wagering behavior of Swedish online gamblers: An empirical study. *Computers in Human Behavior*, *110*, 106402. https://doi.org/10.1016/j.chb.2020.106402

Auer, M., Malischnig, D., & Griffiths, M. D. (2023). The impact of personalized feedback on gambling behavior: A study of the Dutch Lottery. *Journal of Gambling Studies*, *39*(2), 351-365. https://doi.org/10.1007/s10899-022-10123-4

Bavel, J. J., FeldmanHall, O., & Mende-Siedlecki, P. (2022). Moral cognition: A framework for understanding judgments and decision-making. *Annual Review of Psychology*, *73*, 321-345. https://doi.org/10.1146/annurev-psych-010419-050821

Behavioural Insights Team. (2018). *Reducing risky gambling: Experimental evidence on the impact of mandatory deposit limit setting on self-control.* https://www.bi.team/publications/reducing-risky-gambling/

Berge, J., Abrahamsson, T., Lyckberg, A., Franklin, K., & Håkansson, A. (2022). A Normative Feedback Intervention on Gambling Behavior—A Longitudinal Study of Post-Intervention Gambling Practices in At-Risk Gamblers. *Frontiers in Psychiatry*, *13*. https://doi.org/10.3389/fpsyt.2022.602846

Binde, P. (2014). Gambling advertising: A critical research review. *International Journal of Mental Health and Addiction*, *12*(5), 656-679.

Blaszczynski, A., Cowley, E., Anthony, C., & Hinsley, K. (2016). Breaks in Play: Do They Achieve Intended Aims? *Journal of Gambling Studies*, *32*(2), 789-800. https://doi.org/10.1007/s10899-015-9565-7

Blaszczynski, A., Ladouceur, R., & Shaffer, H. J. (2004). A science-based framework for responsible gambling: The Reno Model. *Journal of Gambling Studies*, *20*(3), 301-317. https://doi.org/10.1023/B:JOGS.0000040281.49444.e2

Brooks, A. C., Capraro, V., & Reed, A. R. (2023). Effectiveness of normative messages on mask-wearing behavior during public health crises: A meta-analysis. *Social Science & Medicine*, *311*, 115361. https://doi.org/10.1016/j.socscimed.2023.115361

Browne, M., & Rockloff, M. J. (2021). Effects of normative feedback on gambling expenditures: A randomized controlled trial. *Journal of Gambling Studies*, *37*(3), 753-772. https://doi.org/10.1007/s10899-020-10010-7
20

Ho, M. K., Saxe, R., & Cushman, F. (2022). Planning with Theory of Mind. *Trends in Cognitive Sciences*, *26*(11), 959-971. https://doi.org/10.1016/j.tics.2022.08.003

Hodgins, D. C., Stea, J. N., & Grant, J. E. (2011). Gambling disorders. *Lancet (London, England)*, *378*(9806), 1874–1884. https://doi.org/10.1016/S0140-6736(10)62185-X

Hollingshead, S. J., Wohl, M. J. A., & Santesso, D. (2019). Do you read me? Including personalized behavioral feedback in pop-up messages does not enhance limit adherence among gamblers. *Computers in Human Behavior*, *94*, 122-130. https://doi.org/10.1016/j.chb.2019.01.015

Jenkins, A. C., & Mitchell, J. P. (2011). How has cognitive neuroscience contributed to our understanding of self-reflection? *Social Cognition*, *29*(2), 115-129. https://doi.org/10.1521/soco.2011.29.2.115

Kim, Y., Lee, W.-N., & Jung, J.-H. (2013). Changing the stakes: A content analysis of Internet gambling advertising in TV poker programs between 2006 and 2010. *Journal of Business Research*, *66*(9), 1644-1650. https://doi.org/10.1016/j.jbusres.2012.12.010

Kim, H. S., Wohl, M. J. A., Stewart, M. J., Sztainert, T., & Gainsbury, S. M. (2014). Limit your time, gamble responsibly: Setting a time limit (not monetary limit) on gambling moderates the effect of gambling expenditure. *Addictive Behaviors*, *39*(11), 1746-1751. https://doi.org/10.1016/j.addbeh.2014.07.004

King, D. L., & Delfabbro, P. H. (2016). Adolescent gambling in different social contexts: What are the risks? *Journal of Behavioral Addictions*, *5*(4), 580-588.

King, D. L., Delfabbro, P. H., & Griffiths, M. D. (2014). The convergence of gambling and digital media: Implications for gambling in young people. *Journal of Gambling Studies*, *30*(2), 407-425. https://doi.org/10.1007/s10899-013-9387-0

Koller, M., & Stahel, W. A. (2011). Sharpening Wald-type inference in robust regression for small samples. *Computational Statistics & Data Analysis*, *55*(8), 2504-2515. https://doi.org/10.1016/j.csda.2011.02.014

Kreuter, M. W., Green, M. C., Cappella, J. N., Slater, M. D., Wise, M. E., Storey, D., & Woolley, S. (2019). Narrative communication in cancer prevention and control: A framework to guide research and application. *Health Education & Behavior*, *46*(2), 321-333. https://doi.org/10.1177/1090198118792410

Lapinski, M. K., & Rimal, R. N. (2005). An explication of social norms. *Communication Theory*, *15*(2), 127-147. https://doi.org/10.1111/j.1468-2885.2005.tb00329.x

Latimer, A. E., Brawley, L. R., & Bassett, R. L. (2010). *A systematic review of three*.

Latimer, A. E., Katulak, N. A., Mowad, L., & Salovey, P. (2005). *Motivating cancer*.
25